\documentclass[RNAAS]{aastex}

\begin{document}

\title{Gaia DR3 IDs for TESS Input Catalog Targets}

\author[0000-0003-3702-0382]{Kevin K.\ Hardegree-Ullman}
\affiliation{Caltech/IPAC-NASA Exoplanet Science Institute, 1200 E.\ California Blvd., MC 100-22, Pasadena, CA 91125, USA}
\email{kevinkhu@caltech.edu}

%%%%% ADD AUTHOR INFO HERE %%%%%

\author[0000-0003-4557-1192]{Melanie Swain}
\affiliation{Caltech/IPAC-NASA Exoplanet Science Institute, 1200 E.\ California Blvd., MC 100-22, Pasadena, CA 91125, USA}
\email{mswain@ipac.caltech.edu}

\author[0000-0002-8035-4778]{Jessie L.\ Christiansen}
\affiliation{Caltech/IPAC-NASA Exoplanet Science Institute, 1200 E.\ California Blvd., MC 100-22, Pasadena, CA 91125, USA}
\email{christia@ipac.caltech.edu}

\author[0000-0002-0388-8004]{Emily A.\ Gilbert}
\affiliation{Caltech/IPAC-NASA Exoplanet Science Institute, 1200 E.\ California Blvd., MC 100-22, Pasadena, CA 91125, USA}
\email{egilbert@caltech.edu}

\author[]{Marcy Harbut}
\affiliation{Caltech/IPAC-NASA Exoplanet Science Institute, 1200 E.\ California Blvd., MC 100-22, Pasadena, CA 91125, USA}
\email{mharbut@ipac.caltech.edu}

\author[0000-0002-3239-5989]{Aurora Y.\ Kesseli}
\affiliation{Caltech/IPAC-NASA Exoplanet Science Institute, 1200 E.\ California Blvd., MC 100-22, Pasadena, CA 91125, USA}
\email{aurorak@caltech.edu}

\author[0000-0003-2527-1598]{Michael B.\ Lund}
\affiliation{Caltech/IPAC-NASA Exoplanet Science Institute, 1200 E.\ California Blvd., MC 100-22, Pasadena, CA 91125, USA}
\email{mlund@caltech.edu}

\author[0009-0008-7417-3170]{Meca Lynn}
\affiliation{Caltech/IPAC-NASA Exoplanet Science Institute, 1200 E.\ California Blvd., MC 100-22, Pasadena, CA 91125, USA}
\email{mlynn@ipac.caltech.edu}

\author[0000-0003-2192-5371]{Julian C.\ van Eyken}
\affiliation{Caltech/IPAC-NASA Exoplanet Science Institute, 1200 E.\ California Blvd., MC 100-22, Pasadena, CA 91125, USA}
\email{vaneyken@ipac.caltech.edu}

\begin{abstract}

The TESS Input Catalog (TIC) was built on Gaia Data Release 2 (DR2). To date, there has not been an update to the TIC to incorporate Gaia Data Release 3 (DR3) IDs. In this Research Note, we outline how we cross-matched the TIC with Gaia DR3 IDs, its immediate applications to exoplanet research, and lessons learned when dealing with such large data sets. A full TIC/Gaia DR3 cross-match table is available for download on the Exoplanet Follow-up Observing Program (ExoFOP) website, and a Jupyter Notebook with instructions to reproduce the table is available on GitHub. %Limited to 150 words for RNAAS.

\end{abstract}

%% Keywords should appear after the \end{abstract} command. 
%% The AAS Journals now uses Unified Astronomy Thesaurus concepts:
%% https://astrothesaurus.org
%% You will be asked to selected these concepts during the submission process
%% but this old "keyword" functionality is maintained in case authors want
%% to include these concepts in their preprints.
\section{Introduction} \label{sec:intro}

The Transiting Exoplanet Survey Satellite (TESS) has been surveying the night sky for exoplanets since 2018 \citep{Ricker2015}. The TESS Input Catalog (TIC) was created to help with target selection, and as of v8.0 \citep{Stassun2019,Paegert2021}, was based on the second Gaia data release \citep[DR2;][]{GaiaCollaboration2018}. Gaia DR2 provided a sturdy backbone of $\sim$1.7 billion point sources with measured parallaxes, proper motions, and broadband photometry from which basic stellar properties could be derived. Gaia DR3 provided new and updated measurements for $\sim$1.8 billion point sources \citep{GaiaCollaboration2023}, but the TIC has not been updated to include Gaia DR3 source IDs.

Gaia DR3 represents 34 months of data (compared to 22 months of observations with Gaia DR2), and reflects more precise and accurate measurements than previous data releases due to ongoing improvements in calibrations and data processing \citep{Riello2021}. Since Gaia DR2 and DR3 are independent catalogs \citep[i.e., each data release is a complete reprocessing of the data,][]{Riello2021}, the source IDs between each release do not necessarily directly map to each other. This means there could be multiple DR3 matches for a single DR2 source, especially if multiple sources could now be resolved in DR3 that were not resolved in DR2 \citep{Mora2022}.  \citet{Mora2022} performed a 2\arcsec\ cross-match between Gaia DR3 and DR2, propagating proper motions when available, resulting in a \texttt{dr2\_neighbourhood}\footnote{\url{https://doi.org/10.17876/gaia/dr.3/4}} table with over 2 billion matches. Here, we describe the use of the \texttt{dr2\_neighbourhood} table to perform a cross-match with the TIC to identify Gaia DR3 IDs, some immediate applications of the TIC/Gaia DR3 cross-match table, and implications for upcoming Gaia DR4.

\section{TIC/Gaia DR3 cross-match}

We first downloaded all TIC v8.2 files\footnote{\url{https://archive.stsci.edu/tess/tic\_ctl.html}} and \texttt{dr2\_neighbourhood} files\footnote{\url{https://cdn.gea.esac.esa.int/Gaia/gedr3/auxiliary/dr2\_neighbourhood/}} and compiled the individual files into one large TIC table (1.73 billion entries) and one large Gaia table (2.11 billion entries). The TIC table was limited to the TIC ID and Gaia DR2 ID columns, and entries without a Gaia DR2 ID were removed, leaving us with 1.70 billion TIC targets. \citet{Mora2022} found that 99.56\% of Gaia DR3 sources have one Gaia DR2 match closer than 100 mas, and at this close distance most matches are expected to be the same object, which is supported by most absolute $G$-band magnitude differences being $<0.2$ \citep[see Figure 16.3 of][]{Mora2022}. Therefore, to minimize spurious cross-matches, we limited the Gaia table to targets with a Gaia DR2--DR3 angular distance of $<$100~mas and an absolute Gaia $G$-band magnitude difference of $\le0.2$. These quality cuts left us with 1.67 billion matches. We then merged the TIC and Gaia cross-match tables based on the Gaia DR2 IDs. This resulted in a table of 1.68 billion TIC+Gaia DR3 matches (Table~\ref{tab:match}).

The increase in numbers in that last step is due to there being $\sim$10 million Gaia DR2 (and therefore DR3) targets that correspond to two TIC IDs. This is a known artifact from the update between TIC v7 and TIC v8, and can generally be attributed to an unresolved multiple star in 2MASS that has multiple Gaia DR2 entries. We refer the reader to the TIC v8.2 documentation for more information.\footnote{\url{https://outerspace.stsci.edu/spaces/TESS/pages/111023212/TIC+v8.2+and+CTL+v8.xx+Data+Release+Notes}} We note that $\sim$44 million targets have different Gaia DR2 and Gaia DR3 IDs, and $\sim$1.636 billion targets have the same Gaia DR2 and Gaia DR3 IDs. Within Table~\ref{tab:match}, there are only 73 TIC entries with two Gaia DR3 IDs corresponding to a single Gaia DR2 ID. It is likely that these 73 targets correspond to resolved multiple Gaia DR3 targets that were not resolved in Gaia DR2. We retain each entry, along with the angular distance and $G$-band magnitude difference in Table~\ref{tab:match}. 

\begin{deluxetable}{rcccc}

\tablecaption{TIC/Gaia DR3 cross-match table. Angular distance and G magnitude difference are from the Gaia \texttt{dr2\_neighbourhood} table and are kept here for reference.\label{tab:match}}

\tablehead{\colhead{TIC ID} & \colhead{Gaia DR2 ID} & \colhead{Gaia DR3 ID} & \colhead{Angular Distance} & \colhead{G Magnitude Difference} \\ 
\colhead{} & \colhead{} & \colhead{} & \colhead{(mas)} & \colhead{(mag)}} 

\startdata
1 & 6220232982534277760 & 6220232982534277760 & 0.031861 & -0.012459 \\
2 & 6220232913814800640 & 6220232913814800640 & 0.074949 & -0.027599 \\
3 & 6220232948174541696 & 6220232948174541696 & 0.063944 & -0.010159 \\
4 & 6220232948174542464 & 6220232948174542464 & 0.032790 & -0.017881 \\
5 & 6220233016894020352 & 6220233016894020352 & 0.105402 & -0.010900 \\
7 & 6220233738448530048 & 6220233738448530048 & 0.325959 & -0.025444 \\
8 & 6220233940310564992 & 6220233940310564992 & 0.276488 & -0.012419 \\
9 & 6220233738448529664 & 6220233738448529664 & 0.073262 & -0.015306 \\
10 & 6220234421346907392 & 6220234421346907392 & 0.046712 & -0.018076 \\
11 & 6220233016894022784 & 6220233016894022784 & 0.074753 & -0.010195 \\
\enddata

\tablecomments{This table is available in its entirety on the ExoFOP.}

\end{deluxetable}

\section{Applications}

The primary goal of creating this table was to allow the exoplanet community to use the most recent Gaia information in their research. Both the NASA Exoplanet Archive\footnote{\url{https://exoplanetarchive.ipac.caltech.edu/}} and Exoplanet Follow-up Observing Program (ExoFOP)\footnote{\url{https://exofop.ipac.caltech.edu/}} \citep{Christiansen2025} now include Gaia DR3 IDs for all TIC targets in Table~\ref{tab:match}. For targets with two Gaia DR3 IDs corresponding to a TIC ID, both Gaia DR3 IDs are included as aliases to the TIC ID. A name resolving search tool has been setup on ExoFOP\footnote{\url{https://exofop.ipac.caltech.edu/tess/find_ticid.php}} to allow users to identify TIC, Gaia DR2, and Gaia DR3 IDs for single or multiple targets. The full TIC/Gaia DR3 cross-match table in Table~\ref{tab:match} is available for download on the ExoFOP.\footnote{\url{https://exofop.ipac.caltech.edu/tess/gaiadr3_tic_crossmatch.php}} There are also several large stellar spectroscopic databases (e.g., APOGEE, DESI MWS, GALAH, LAMOST) that include Gaia DR3 IDs in their catalog, for which the TIC/Gaia DR3 cross-match table could simplify TESS target identification.

\section{Preparing for Gaia DR4 and lessons learned}

Although the Gaia spacecraft collected its last photons in January 2025, it will take several more years for all the data to become available. Gaia DR4 is currently expected to release in December 2026.\footnote{\url{https://www.cosmos.esa.int/web/gaia/release}} Assuming Gaia DR4 provides similar cross-match tables (DR4--DR3 or DR4--DR2), we can use the same methodology outlined in this research note to quickly provide updated Gaia DR4 IDs to TIC targets.

It is worth noting that handling such large data sets ($\sim$50--125~GB) came with much trial-and-error. This analysis was performed on a standard laptop (Intel Core i7-11800 processor, 64~GB RAM) running Linux (Ubuntu), but 300~GB of swap memory was necessary to perform all table functions. After several unsuccessful attempts at merging the TIC and Gaia tables, we found the \texttt{Polars} library \citep{Vink2025} was able to merge the two tables in about 4 minutes.\footnote{Prior to merging, we recommend sorting each table by the column on which you intend to merge.} For full reproducibility of this table, we provide a Jupyter Notebook on GitHub.\footnote{\url{https://github.com/kevinkhu/ticgaia}}

%% Please use the acknowledgment and contribution environments. This will 
%% be anonomyized when the "anonymous" style option is used. 
\begin{acknowledgments}
Some of the data presented in this paper were obtained from the Mikulski Archive for Space Telescopes (MAST) at the Space Telescope Science Institute. The specific observations analyzed can be accessed via \dataset[https://doi.org/10.17909/fwdt-2x66]{https://doi.org/10.17909/fwdt-2x66}. STScI is operated by the Association of Universities for Research in Astronomy, Inc., under NASA contract NAS5–26555. Support to MAST for these data is provided by the NASA Office of Space Science via grant NAG5–7584 and by other grants and contracts.

This work has made use of data from the European Space Agency (ESA) mission Gaia (\url{https://www.cosmos.esa.int/gaia}), processed by the Gaia Data Processing and Analysis Consortium (DPAC, \url{https://www.cosmos.esa.int/web/gaia/dpac/consortium}). Funding for the DPAC has been provided by national institutions, in particular the institutions participating in the Gaia Multilateral Agreement.

This research has made use of the NASA Exoplanet Archive, which is operated by the California Institute of Technology, under contract with the National Aeronautics and Space Administration under the Exoplanet Exploration Program.

This research has made use of the Exoplanet Follow-up Observation Program (ExoFOP; DOI: 10.26134/ExoFOP5) website, which is operated by the California Institute of Technology, under contract with the National Aeronautics and Space Administration under the Exoplanet Exploration Program.

\end{acknowledgments}

% \begin{contribution}
%%This section gives authors the space to recognize author contributions. The text inside this environment is NOT counted towards the total word quanta. At a minimum, manuscripts are expected to include this text:

% All authors contributed equally to the Terra Mater collaboration.

%% But authors are expected to provide more specific details, e.g. 
%%
%%SC was responsible for writing and submitting the manuscript.
%%WWM came up with the initial research concept and edited the manuscript.
%%OTS obtained the funding and edited the manuscript.
%%EBF provided the formal analysis and validation. He also edited the manuscript.
%%GEH Supervised the undergraduates, wrote the software and administers the project github and Zenodo repositories.
%%
%% Authors can use the Contributor Role Taxonomy (CRediT) at
%% https://credit.niso.org
%% for ideas on how write a good statement tailored to their needs.

% \end{contribution}

%% To help institutions obtain information on the effectiveness of their 
%% telescopes the AAS Journals has created a group of keywords for telescope 
%% facilities.
%
%% Following the acknowledgments section, use the following syntax and the
%% \facility{} or \facilities{} macros to list the keywords of facilities used 
%% in the research for the paper.  Each keyword is check against the master 
%% list during copy editing.  Individual instruments can be provided in 
%% parentheses, after the keyword, but they are not verified.
\facilities{ExoFOP, Exoplanet Archive, Gaia, TESS}

\software{\texttt{Jupyter} \citep{Kluyver2016}, \texttt{Polars} \citep{Vink2025}}

\newpage
\bibliography{main}{}
\bibliographystyle{aasjournalv7}

\end{document}